\documentclass[prb,twocolumn,superscriptaddress,showpacs,floatfix, longbibliography]{revtex4-1}%

\usepackage{amsmath,amssymb}
\usepackage{graphicx}
\usepackage{dcolumn}
\usepackage{soul}
\usepackage{xcolor}
\definecolor{darkgreen}{rgb}{0.1, 0.5, 0.16}
\begin{document}


\title{Superconducting Chevrel phase PbMo$_{6}$S$_{8}$ from first principles} 

\author{Giovanni Marini}
\affiliation{Department of Physical and Chemical Sciences,  University of L'Aquila, Via Vetoio 10, I-67100 L'Aquila Italy}
\affiliation{SPIN-CNR, University of L'Aquila, Via Vetoio 10, I-67100 L'Aquila Italy}

\author{Antonio Sanna} 
\affiliation{Max-Planck Institut f\"ur MikrostrukturPhysik, Weinberg 2, 06120 Halle, Germany}

\author{Camilla Pellegrini} 
\affiliation{Department of Physics and Geology, University of Perugia, Via Pascoli 33, 06123 Perugia, Italy}

\author{Christophe Bersier}
\affiliation{Max-Planck Institut f\"ur MikrostrukturPhysik, Weinberg 2, 06120 Halle, (Germany)}

\author{Erio Tosatti}
\affiliation{International School for Advanced Studies (SISSA), Via Bonomea 265, 34136 Trieste, Italy}
\affiliation{CNR-IOM Democritos National Simulation Center, Via Bonomea 265, 34136 Trieste, Italy}
\affiliation{The Abdus Salam International Centre for Theoretical Physics (ICTP), Strada Costiera 11, 34151 Trieste, Italy}

\author{Gianni Profeta}
\affiliation{Department of Physical and Chemical Sciences,  University of L'Aquila, Via Vetoio 10, I-67100 L'Aquila (Italy)}
\affiliation{SPIN-CNR, University of L'Aquila, Via Vetoio 10, I-67100 L'Aquila (Italy)}

\begin{abstract}
Chevrel ternary superconductors show an intriguing coexistence of molecular aspects, large electron-phonon and electron-electron correlations, which to some extent still impedes their quantitative understanding. 
We present a first principles study on the prototypical Chevrel compound PbMo$_{6}$S$_{8}$, including electronic, structural and vibrational properties at zero and high pressure. 
We confirm the presence of an extremely strong electron-phonon coupling, linked to the proximity to a R$\overline{3}$-P$\overline{1}$ structural phase transition, which weakens as the system, upon applied pressures, is driven away from the phase boundary.  
A detailed description of the superconducting state is obtained by means of fully \textit{ab initio} superconducting density functional theory (SCDFT). 
SCDFT accounts for the role of phase instability, electron-phonon coupling with different intra- and inter-molecular phonon modes, and without any empirical parameter, and accurately reproduces the 
experimental critical temperature and gap. This study 
provides the conclusive confirmation
that Chevrel phases are phonon driven superconductors 
mitigated, however, by
an uncommonly strong Coulomb repulsion. The latter is generated by the combined effect of repulsive Mo states at the Fermi energy and a band gap in close proximity to the Fermi level. This is crucial to rationalize why Chevrel phases, in spite of their extreme electron-phonon coupling, have critical temperatures below 15~K.
In addition, we predict the evolution of the superconducting critical temperature as a function of the external pressure, showing an excellent agreement with available experimental data. 
\end{abstract}

\maketitle

\section{\label{sec:level1}Introduction}
Molybdenum chalcogenides, better known as Chevrel phases, were discovered in 1971 by Roger Chevrel and collaborators
\cite{Chevrel1971} as the first known ternary superconductors (SC)\cite{Matthias1972}. 
To date, this family counts more than one hundred examples, and has been extensively investigated~\cite{Pea2015}. 
The general stoichiometry of Chevrel phases is M$_{y}$Mo$_{6}$X$_{8}$ ( M = metal and X = chalcogen), but they can often deviate from the stoichiometric formulae in experimental observations. Their peculiar crystal structure, characterized by molecular-like interpenetrating Mo$_{6}$ octahedra and X$_{8}$ cubes, allows for a wide chemical variability as different chemical species can be intercalated into the structure, a fact which is at the origin of the broad spectrum of interesting physical properties they display. Owing to the peculiar cage-like structure, Chevrel phases have been considered as promising thermoelectric materials\cite{Rao1984,Caillat1999,Roche1998,Nunes1999}, as cathode materials in battery design\cite{Uchida1990,Suzuki1998} and, when X=S, as catalysts in desulfurization processes\cite{Benson1995,Kamiguchi2012,Afanasiev2007}.
As superconductors, the importance of Chevrel phases is linked to the extremely high critical magnetic field (H${_c}_2$), up to $\approx$ 60 Tesla\cite{Fischer1978}, that combines with critical temperatures (T$_{c}$) as high as 15~K. For these figures Chevrels have been considered a valuable alternative to the A15 SC class for the fabrication of SC magnets~\cite{Kubo1987,Seeber1989}. 

From a fundamental point of view, this family represents a unique platform to study the interplay between superconductivity and magnetism, when the cations are magnetic atoms. For example, REMo$_{6}$S$_{8}$  and REMo$_{6}$Se$_{8}$ (RE = rare-earth) show coexistence between superconductivity and long-range magnetic ordering\cite{McCallum1977}. Moreover, re-entrant SC transitions and ferromagnetic phases have been reported for ErRh$_{4}$B$_{4}$ and HoMo$_{6}$S$_{8}$ (many details on the subject can be found in the reviews of Pe\~{n}a\cite{Pea2015} and Fischer\cite{Fischer1978}).

Research on Chevrel phases has thrived in the 70s and 80s, until the discovery of cuprate superconductors~\cite{BednorzMueller_BaLaCuO_Zphys1986} shifted the attention away, leaving behind an incomplete characterization with several unsolved problems, especially concerning the nature and properties of the superconducting pairing. 
In the Uemura plot\cite{Uemura1989,Hillier1997}, which relates the SC critical temperature T$_{c}$ to the ratio between carrier density and effective mass, the Chevrel phases lay in the region of unconventional superconductors, together with cuprates, iron-based superconductors, fullerenes, organics superconductors and heavy-fermions, and far from the ``conventional'' BCS-like part of the graph. Nevertheless, the origin of the superconducting transition in Chevrel phases is attributed to the electron-phonon mechanism\cite{Fischer1978,Bader1978,Reichman}. In addition, a recent experiment reports the appearance of a double SC gap in PbMo$_{6}$S$_{8}$, SnMo$_{6}$S$_{8}$ and AgMo$_{6}$S$_{8}$ \cite{Petrovi2011,Feig2017}. Attempts to clarify the situation with modern theoretical methods have been scarce, and few works provide \textit{ab initio} results on the topic\cite{Roche1998,Reichman,Nunes1999}. 
Only recently, Chen {\em et al.}\cite{Reichman} afforded a first-principles density functional theory description of Chevrel phases. They predicted an electron-phonon coupling (EPC) parameter, $\lambda$, (2.29 for PbMo$_{6}$S$_{8}$), much larger than previous estimates\cite{Fischer1978} blue predicting, in fact, SC critical temperatures systematically higher than the experimental values reported for (Pb, Sn, Yb, La, Y) Mo$_{6}$S$_{8}$\cite{Matthias1972,Shelton1975}, therefore still leaving questions on the origin and quantitative understanding of the superconducting transition.

The aim of the present work is to definitely shed light on these unsolved issues in the characterization of Chevrel phases by reinvestigating their electronic and vibrational properties, and by giving an account of their superconducting state using density functional theory for superconductors (SCDFT)\cite{OGK_SCDFT_PRL1988, Lueders2005,Marques2005}, an \emph{ab-initio} cutting-edge approach to phonon-mediated superconductivity. Due to the accuracy of the method, a careful comparison of our predictions with available experimental data at ambient and high pressures~\cite{Shelton1975} is expected to detect anomalies in the superconducting behaviour of Chevrel phases. We focus our attention on the PbMo$_{6}$S$_{8}$ compound, as the prototype system with highest T$_c$ of a wide class of Chevrel phases,
therefore extending and complementing previous theoretical approaches, particularly the latest by Chen {\em et al.}~\cite{Reichman}.
Incidentally, we note that a tempting analogy could initially have been drawn with alkali fullerides (doped molecular crystals with high $T_c$~\cite{Hebard_KC60_Nature1991}, strong electron-phonon couplings \cite{Gunnarson_SCFullerides_RMP1997} and very strong electron-electron correlations \cite{LofSawatzky_BandGapExcitonsCoulombC60_JESRP1995}). For them it was proposed\cite {Capone_StrCorrSC_Science2002, Capone_ModellingUnconventionalSC_RMP2009}, and experimentally demonstrated, particularly for $Cs_3C_{60}$  \cite{Takabayashi_UnconvSCFullerides_PhilTranRSA2016}, that superconductivity is strongly correlated, arising from a  low-spin Mott insulator \cite{Fabrizio_NonMagJTMottIns_PRB1997} after doping and metallization.
We will show, first of all, that negative pressures on Chevrel phases can lead to electron-phonon driven structural transitions. However, they do not lead to metal-insulator or magnetic states, that might suggest neighborhood to Mott insulators or strongly correlated superconductivity mechanisms analogous to those of alkali fullerides \cite{Capone_ModellingUnconventionalSC_RMP2009}. 
This justifies the treatment of Chevrels as phononic superconductors. Nonetheless, Coulomb interactions are extremely important, in particular their energy dependence afforded by SCDFT, and the main result of our  first-principles treatment will be to explain for the first time the anomalously low critical temperature of PbMo$_{6}$S$_{8}$, attained despite the large electron-phonon coupling.

\section{Methods}
\label{sec:methods}
First-principles calculations of normal state properties have been performed within  density functional theory (DFT). We have used norm-conserving pseudopotentials generated from the Troullier-Martins scheme\cite{Troullier1991} to describe the electron-ion interaction, while employing a kinetic energy cutoff of 60~Ry in the plane-wave expansion of the Kohn-Sham wavefunctions, as implemented in the \textit{Quantum Espresso} package \cite{ref7}. For the exchange-correlation potential we have adopted the generalized gradient approximation (GGA) in the Perdew, Burke and Ernzerhof (PBE)\cite{ref8} formulation. A 4$\times$4$\times$4 Monkhorst-Pack wave-vector grid\cite{ref15} has been required to reach an accuracy in the total energy better than 1 meV/atom, while the energy cutoff has been increased to 80 Ry in order to converge the stress tensor in the variable-cell calculations. The structural phase diagram has been constructed by performing a variable-cell optimization of the structures at fixed pressures, and then calculating the corresponding enthalpy ($H(P)=E(V(P))+PV(P)$). 

The lattice dynamics has been simulated using density functional perturbation theory (DFPT)\cite{PhysRevLett.58.1861}, with a $8^3$ ($4^3$) sampling 
in \textbf{k} (\textbf{q}) space of the electronic (phononic) wavefunctions in the Brillouin Zone (BZ). 
Accurate integration of the electron-phonon matrix elements $g_{\textbf{k}+\textbf{q},\textbf{k}}^{\nu, m n}$ in the Eliashberg function:
\begin{eqnarray} 
 \alpha^2F(\omega) &=& \frac{1}{N(E_{F})} \sum\limits_{{\bf k}{\bf q}\nu}\sum_{n m}  |g_{\textbf{k}+\textbf{q},\textbf{k}}^{\nu, m n}|^2 \nonumber \\ &\times& \delta(\omega-\omega_{{\bf q}\nu})\delta(\epsilon_{\textbf{k}+\textbf{q}}^m - E_{F}) \delta(\epsilon_{\textbf{k}}^n - E_{F}),  \label{eq:alpha}
\end{eqnarray}
has been achieved by using a random mesh of \textbf{k} points accumulated on the Fermi surface by means of a Metropolis algorithm, and by Fourier interpolating the calculated phonon frequencies  ($\omega_{{\bf q}\nu}$)  and electronic eigenvalues 
$(\epsilon_{\textbf{k}}^n$~\cite{Marques2005,Floris2005,Floris2007}. The integrated EPC  constant $\lambda$ has been obtained as:
\begin{eqnarray} 
 \lambda &=& 2 \int \frac{\alpha^2F(\omega)}{\omega}d\omega. \label{eq:lambda}
\end{eqnarray}

To compare with conventional theoretical approaches, we have estimated the superconducting critical temperature using the Allen-Dynes-modified McMillan formula
\begin{equation}  T_{c} = \frac{f_{1}f_{2}\omega_{ln}}{1.20} {\rm exp}\left[\frac{-1.04(1+\lambda)}{\lambda-\mu^{*}(1+0.62\lambda)}\right] \label{eq:AllenDynesTc} \end{equation}
where $\mu^{*}$ is the renormalized Coulomb pseudopotential (see below), $\omega_{ln}$ is the logarithmic average phonon frequency and $f_{1}$,$f_{2}$ are correction factors introduced in Ref.~\onlinecite{AllenDynes_PRB1975}. The parameter $\mu^*$ describes the effective strength of the residual Coulomb interaction within the Cooper pairs. It is often used as an adjustable parameter but its 
``conventionally" accepted value ranges between 0.1 for $sp$ metals and 0.15 for transition metals~\cite{AllenMitrovic_TheoryofSuperconductingTc_1983}. Using  Morel-Anderson theory~\cite{MorelAnderson_PR1962,ScalapinoSchriefferWilkins_StongCouplingSC_PR1966}, $\mu^*$ can be linked to 
the Coulomb potential 
\begin{equation} 
\mu = N(E_{F}) V_{c}(E_F, E_F),\label{mu} \end{equation}
where $E_F$ is the Fermi energy, 
$N(\epsilon)$ is the electronic density of states (DOS) at energy $\epsilon$, and $V_c(\epsilon,\epsilon')$ is the screened Coulomb interaction~\cite{Sanna_GenuinePredictions_JPSJ2018,Davydov_AbinitioEliashberg_arxiv2020,Massidda_SUST_CoulombSCDFT_2009,Sanna_CaC6_PRB2007}. The latter is defined as an iso-energy surface average:
\begin{equation}
V_{c}\left(\epsilon,\epsilon'\right)=\frac{1}{N\left(\epsilon\right) N\left(\epsilon'\right) }\sum_{n{\bf k},n{\bf k}'}V^{n,n'}_{c\,{\bf k}{\bf k}'} \delta\left(\epsilon-\epsilon^n_{\bf k}\right)\delta(\epsilon'-\epsilon^{n'}_{{\bf k}'}),\label{eq:coulomb}
\end{equation}
of the screened Coulomb matrix elements: 
\begin{equation}
V^{n,n'}_{c\,{\bf k}{\bf k}'}  =  \frac{4\pi}{\Omega}\sum\limits _{\mathbf{GG}'}\epsilon_{\mathbf{G}\mathbf{G}'}^{-1}\left(\mathbf{q},0\right)
   \frac{\rho^{nn'}_{{\bf k}{\bf k}'}\left(\mathbf{G}\right)\rho^{nn'*}_{{\bf k}{\bf k}'}\left(\mathbf{G}'\right)}{\left|\mathbf{q}+\mathbf{G}\right|\left|\mathbf{q}+\mathbf{G}'\right|},\label{eq:coul_matr_el}
\end{equation}
where $n$ is the band index, $\Omega$ denotes the unit cell volume, $\mathbf{G}$ are reciprocal lattice vectors, ${\bf q}={\bf k}-{\bf k}'$, $\rho^{nn'}_{{\bf k}{\bf k}'}\left(\mathbf{G}\right)=\left\langle n{\bf k}|e^{i\left(\mathbf{q}+\mathbf{G}\right)\mathbf{r}}|n'{\bf k}'\right\rangle$ represents the polarization matrix and $\epsilon_{\mathbf{G}\mathbf{G}'}^{-1}\left(\mathbf{q},\omega\right)$  is the inverse dielectric function that, in this work, is computed within the static random phase approximation (RPA) as implemented in the Elk code\cite{ElkCode}.

A fully \textit{ab initio} alternative to the McMillan approach is superconducting density functional theory~\cite{OGK_SCDFT_PRL1988}, where both the $\alpha^2F$ function and $V_c(\epsilon,\epsilon')$ enter on equal footing. 
SCDFT is an extension of standard DFT for \emph{ab initio} calculations of material-specific
properties in the superconducting state~\cite{OGK_SCDFT_PRL1988}.
The SCDFT approach has proven to provide reliable predictions of the experimental T$_c$'s for many conventional superconductors~\cite{Floris2005,Floris2007,Sanna_CaC6_PRB2007,Cudazzo2008,Profeta_LiKAl_PRL2006,Akashi_HfLiMNCl_PRB2012,Akashi_dopedFullerene_PRB2013,Livas_SH3_EPJB2016,Akashi_SH3_PRB2015,Livas_PunderPress_PRM2017,Arita_NonEmpiricalLightElement_AdvMat2017,FloresLivas_PH_PRB2016,Pellegrini_WO3_PRM2019}. 
Within a Kohn-Sham (KS) scheme, the central equation to be solved in SCDFT is a BCS-like gap equation~\cite{Lueders2005,Marques2005,Davydov_AbinitioEliashberg_arxiv2020,SPG_EliashbergSCDFT_PRL2020}, which in the isotropic limit reads as:
\begin{equation}
\label{eq:SCDFTgapeq}
\Delta_{s,\epsilon}=-\mathcal{Z}_{\epsilon}\Delta_{s,\epsilon}-\frac{1}{2}\int d\epsilon'\mathcal{K}_{\epsilon,\epsilon'}\frac{\tanh\left(\frac{\beta}{2}E_{\epsilon'}\right)}{E_{\epsilon'}}\Delta_{s,\epsilon'},
\end{equation}
 where $\Delta_{s,\epsilon}$ is the KS superconducting gap, 
$E_{\epsilon}=\sqrt{\epsilon^2+|\Delta_{s,\epsilon}|^2}$ represent the KS excitation energies and $\beta$ is the inverse temperature.
The exchange-correlation kernels to be used in the equation are $\mathcal{Z}_{\epsilon}=\mathcal{Z}^{ph}_{\epsilon}$ and  $\mathcal{K}_{\epsilon,\epsilon'}=\mathcal{K}^{ph}_{\epsilon,\epsilon'}+\mathcal{K}^c_{\epsilon,\epsilon'}$. 

Here, the Coulomb term is simply given by
\begin{equation}
   \mathcal{K}^c_{\epsilon,\epsilon'}=N(\epsilon')V_{c}\left(\epsilon,\epsilon'\right).\label{eq:Kc}
\end{equation}
As for the phononic kernels, $\mathcal{Z}^{ph}_{\epsilon}$ and $\mathcal{K}^{ph}_{\epsilon,\epsilon'}$, we have used the recently developed SPG functional~\cite{SPG_EliashbergSCDFT_PRL2020}, which, having been constructed from the Eliashberg self-energy, allows for accurate estimations of both the T$_c$ and the superconducting gap function.

\section{Results}\label{results}
\subsection{Phase diagram and equilibrium structure}\label{sec:phasediagram}

At ambient conditions Chevrel phases typically crystallize in the rhombohedral R$\overline{3}$ lattice (space group  148), although some members of the family lower their symmetry by a triclinic distortion, ending up in the P$\overline{1}$ symmetry group (space group 2). Fischer classified Chevrel phases into two main classes\cite{Fischer1978}. In the compounds of the first class, the cation M has a small ionic radius and its concentration $y$ can vary across a continuous range, as is the case for M = Cu, Ti, Cr. In the second class of materials, the element M is a large cation, e.g., a lanthanide, alkaline earth, Pb or Sn, and its concentration $y$ is fixed (or has a narrow variation range due to defects) with maximum occupancy of one atom per unit cell.

\begin{figure}
\centering
\includegraphics[width=1\linewidth]{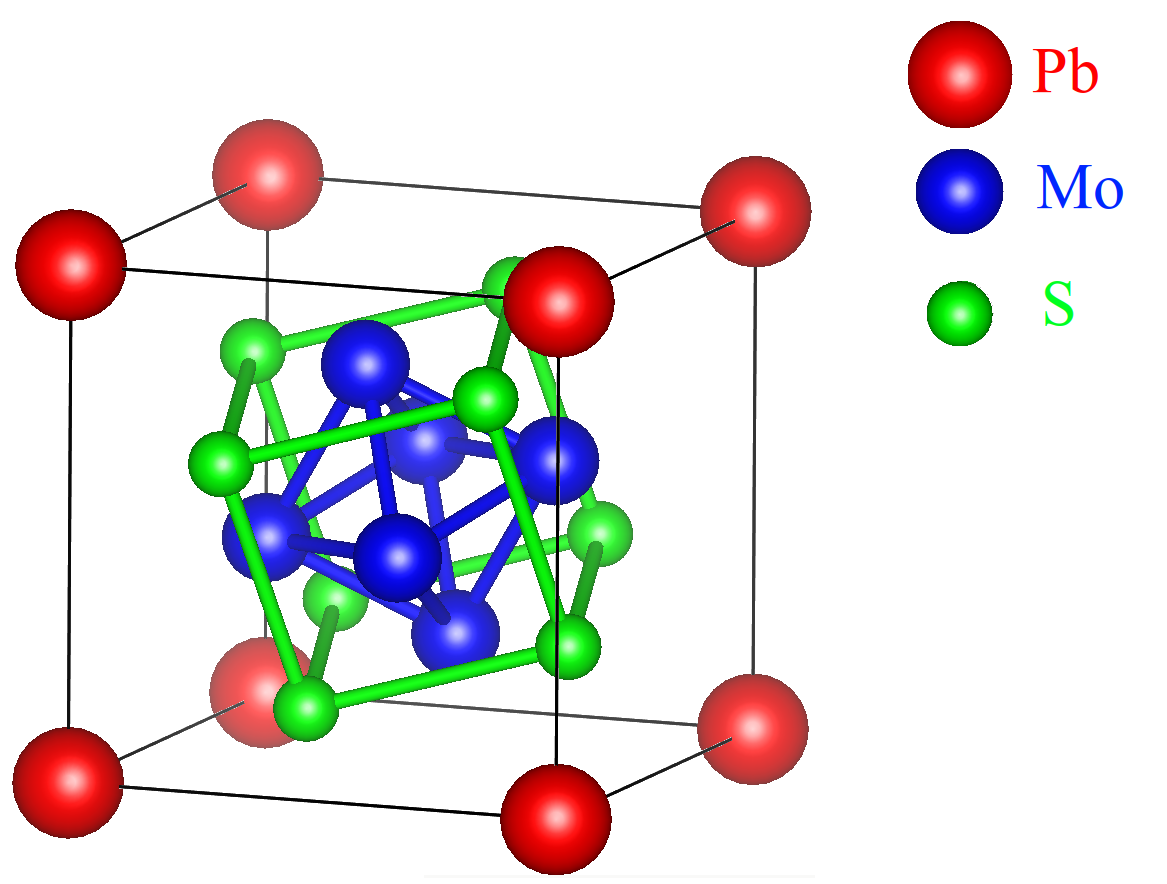}
\caption{Crystal structure of the Chevrel phase PbMo$_{6}$S$_{8}$ in the R$\overline{3}$ phase\cite{ref14}.}\label{fig1}
\end{figure}

The prototype crystal structure for the PbMo$_{6}$S$_{8}$ compound is sketched in Fig.\ref{fig1}. 
Here, the metal atom, M=Pb, occupies the 3a $(0, 0, 0)$ Wyckoff position of the rhombohedral R$\overline{3}$ unit cell. However, smaller cations can occupy either the 3a or the 18f Wyckoff positions up to a maximum of four cations per formula unit. The main structural unit  consists of 6 molybdenum atoms caged by a quasi-cubic structure of 8 chalcogen atoms. These ``molecular'' units are arranged in a closely packed structure and are rotated by an angle $\theta \simeq 25^{\circ}$ with respect to the threefold rhombohedral axis, in such a way to favour inter-molecular bonds between neighbouring units. 

Experimentally, at pressure P=0, PbMo$_{6}$S$_{8}$ has  R$\overline{3}$ crystal structure~\cite{Chevrel1971},   superconducting T$_{c}$ of 15~K\cite{Marezio1973,Shelton1975} and critical field of $\approx$50~ Tesla\cite{Fischer1978}. 

For the R$\overline{3}$ structure of PbMo$_{6}$S$_{8}$, the equilibrium structural parameters obtained using the PBE functional are $a$= 6.60 \AA\ with a rhombohedral angle of 89.32$^\circ$. These estimates are in good agreement with the experimental values of 6.55 \AA\ and 89.33$^\circ$\cite{Marezio1973}. However, as expected in PBE\cite{Grabowski2007},  we have a slight
overestimation of the unit cell volume,
 roughly equivalent to a negative pressure of -2.1GPa applied to the experimental cell.

\begin{figure}
\centering
\includegraphics[width=1\linewidth]{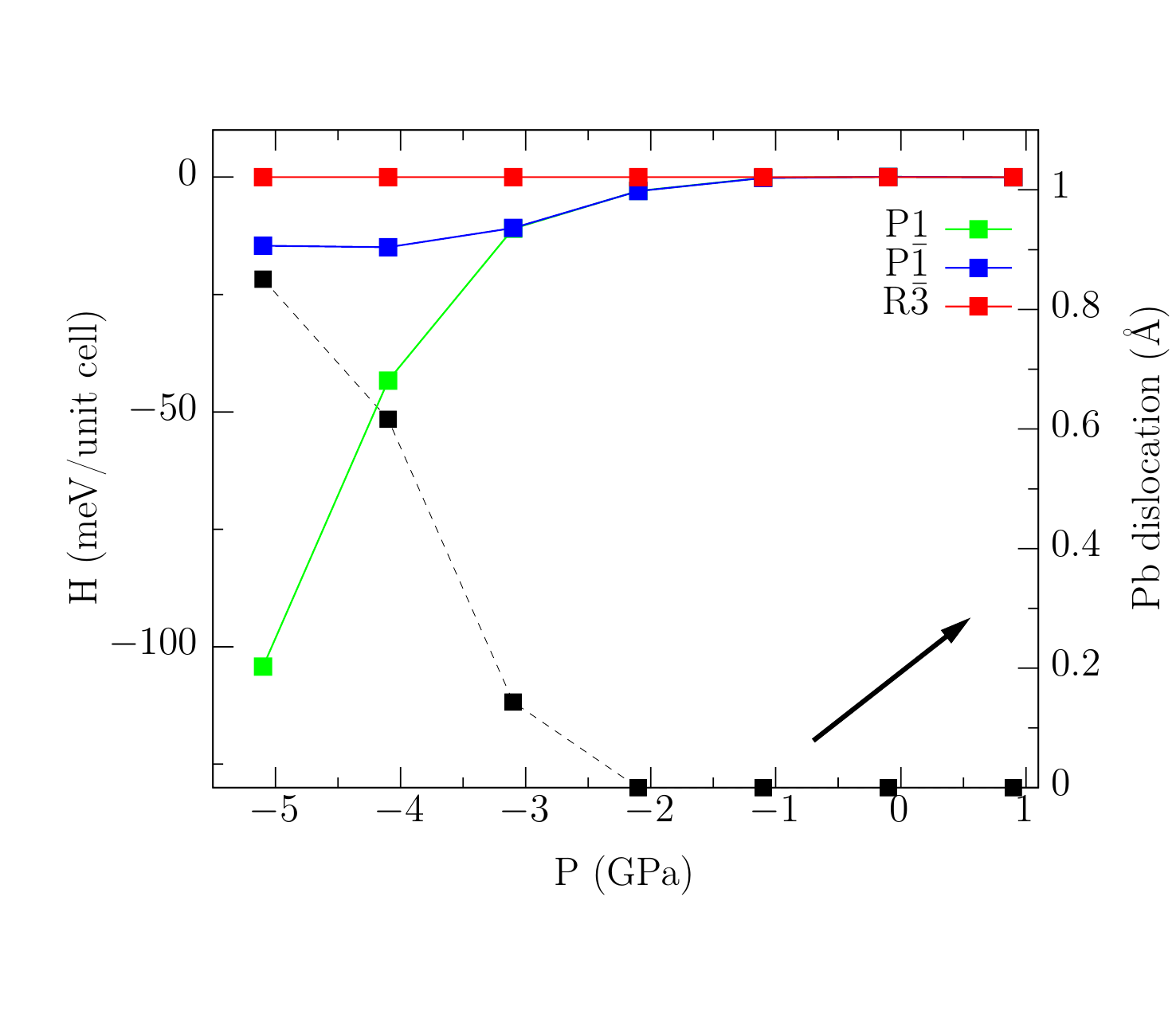}

\vspace{-1cm}

\caption{Enthalpy-pressure phase diagram for PbMo$_{6}$S$_{8}$ at T = 0 K, referred to the enthalpy of the R$\overline{3}$ crystal phase (red). We found a triclinic (R$\overline{3}\to$P$\overline{1}$) distortion at $\sim$ -1 GPa, and (P$\overline{1}\to$P1) transition at about -3 GPa losing inversion symmetry, due to a Pb shift from the origin in the P1 crystal phase. The amount of the displacement is shown by a black dashed line (which refers to the right axis).}\label{fig:Enthalpy}
\end{figure}

This trigonal R$\overline{3}$ structure turns out to be energetically in competition with two triclinic structural phases, which are also observed in the Chevrel family\cite{Chevrel1971,Yvon1978,Jorgensen1987}: the P$\overline{1}$ (inversion symmetric) and the P1 (space group 1 - with no inversion symmetry). 
PbMo$_6$S$_8$ at P=0 is, essentially, on the verge of a structural phase transition\cite{Fischer1978} which, owing to the volume overestimation in PBE, is located at a slight positive, rather than negative, pressure. In order to be consistent with experiments, we have applied a rigid shift of +2.1 GPa to the enthalpy-pressure diagram so to match the theoretical and experimental values of the equilibrium volume, thus avoiding  spurious effects in the phonon and electron-phonon calculations. This upward shift in pression, applied once for all, will be implicitly assumed throughout the rest of the paper. The resulting enthalpy-pressure (H(P)) phase diagram at zero temperature, including the three competing phases, is shown in Fig.~\ref{fig:Enthalpy}. 
We did not find  in the whole range, including  negative pressures, any evidence of metal-insulator, magnetic or antiferromagnetic transitions, that might have suggested the neighborhood of  Mott insulating states -- a result that, together with the extremely strong electron-phonon coupling, supports a conventional phononic treatment of this system.  

Despite the small energy differences involved, we successfully predict the R$\overline{3}$ structure as the  stable phase for positive pressures. At slight negative pressures the P$\overline{1}$ triclinic distortion becomes energetically favored (although with a small stabilization energy of only 3 meV/unit cell at -2 GPa). Below -3 GPa there is a clear indication of phase transition towards P1. The breaking of spatial inversion symmetry accompanying the P1 distortion is signaled by the displacement of the Pb atom from the origin of the unit cell (black line in Fig.~\ref{fig:Enthalpy}).

The obtained H(P) phase diagram can be compared to the qualitative phase diagram for Chevrel phases introduced by Jorgensen {\em et al.} in Ref.~\onlinecite{Jorgensen1987}. Here it is shown that different Chevrel sulfur compounds are described by the same pressure-temperature phase diagram when the zero of the pressure axis is shifted according to the chemical pressure induced in each compound by the cation M. In particular PbMo$_{6}$S$_{8}$ at 0 K is reported to fall in a region of coexistence between the R$\overline{3}$ and P$\overline{1}$ phases. Furthermore, Yvon {\em et al.}\cite{Yvon1978} reported the tendency of Pb to slightly shift from the centrosymmetric site, thus breaking inversion symmetry. 

The experimental observation of all the three phases (R$\overline{3}$, P$\overline{1}$ and P1) can be explained by their quasi-degeneracy. Defects and non-ideal stoichiometry, which are in fact ubiquitous in Chevrel phase samples \footnote{The actual stoichiometric formula of Chevrel phase samples has been reported to be Pb$_{0.92}$Mo$_{6}$S$_{8}$, Pb$_{0.92}$Mo$_{6}$S$_{7.5}$, PbMo$_{6.35}$S$_{8}$ in various instances, see e.g. Ref.~\onlinecite{Fischer1978}.}, give rise to  pressure and/or   composition inhomogeneities, which eventually stabilize domains of both the symmetric (R$\overline{3}$) and the distorted phases. This scenario is validated by our H(P) phase diagram which shows three nearly degenerate (or coexisting) structures in a pressure interval of 2-3 GPa around P=0.

\subsection{Electronic properties}\label{sec:electrons}

\begin{figure}
\centering
\includegraphics[width=1\linewidth]{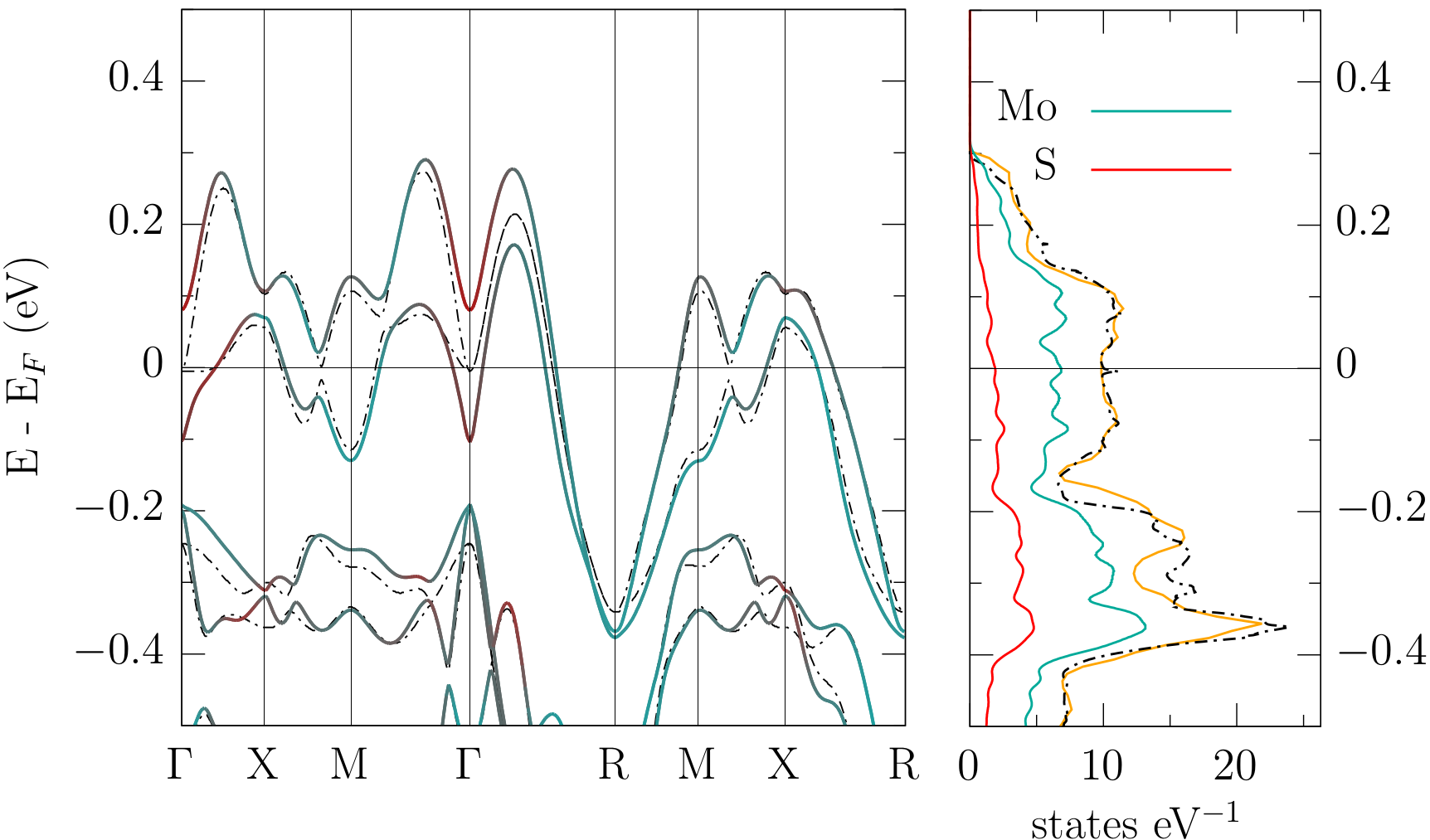}
\caption{Kohn-Sham electronic states for the equilibrium structure with SOC (projected over atomic Mo $d$-states (teal) and S $p$-states (red)) and without SOC (black dashed lines). Side panel: total DOS with SOC (orange solid line), without SOC (black dashed line)  and projected over Mo $d$-states and S $p$-states (teal and read solid lines respectively). }\label{fig:elBANDSandDOS}
\end{figure}

The atom-projected band structure and DOS of PbMo$_6$S$_8$ are shown in Fig.~\ref{fig:elBANDSandDOS}. Despite the complex geometry, the chemical bonding can be described in terms of a simple model\cite{YVON197741}, where the cation M transfers its nominal charge to the Mo$_{6}$X$_{8}$ cluster. Following Ref.~\onlinecite{YVON197741}, the valence electron concentration (VEC) of the generic Chevrel phase M$_{x}$Mo$_{6}$X$_{8}$ can be calculated as follows
\begin{align}
VEC &= 6\times6(e_{Mo}) - 2\times8(e_X) + x \times e_{M} \nonumber \\ 
    &= 20 + x \times e_{M},
\end{align}
where $e_A$ indicates the valence-electron contribution of the element A. In Chevrel phases, the VEC can vary from 20 to 24. 
For PbMo$_{6}$S$_{8}$ the VEC is 22, meaning that the Fermi level lies at half filling of the E$_{g}$ bands. Indeed, this simple picture is confirmed by DFT calculations of the band structure. 
In the energy window from -2 eV to 3 eV all the states originate from the hybridization between chalcogens $p$-states and molybdenum $d$-states. On the contrary, in the valence region between -2 eV and -5 eV, one finds a predominance of chalcogen $p$-states with a minor Mo content. As shown in Fig.~\ref{fig:elBANDSandDOS}, the Fermi level is crossed by two doubly-degenerate narrow ($\approx$ 0.5 eV) E$_{g}$ bands with strong $d(x^2-y^2)$ character\cite{Andersen_ElectronicStructureChevrel_PRB1978}. These can accommodate up to four electrons per unit cell, resulting in their half-filling. 
We mention that, due to the small trigonal distortion of the unit cell of PbMo$_{6}$S$_{8}$ (see Fig.~\ref{fig1}), the band structure in Fig.~\ref{fig:elBANDSandDOS} has been calculated for a cubic BZ instead of using the (more complicated) trigonal BZ.

Like in its elemental {\em bcc} phase\cite{spin-orbit-Mo}, the molybdenum ion in PbMo$_{6}$S$_{8}$ has strong spin-orbit coupling (SOC). As one can see in Fig.~\ref{fig:elBANDSandDOS}, the inclusion of the SOC modifies the band dispersion by splitting the $\Gamma$-$R$ line and producing a small Fermi surface (FS) pocket around the $\Gamma$ point. We are not aware of experimental investigations of the FS topology of Chevrel phases, therefore we call for dedicated experiments to measure the PbMo$_6$S$_8$ FS or band structure. However, the SOC contribution does not alter the shape of the DOS close to the Fermi level, or its value at the Fermi energy, N(E$_{F}$) (which changes from 10.1 to 9.9 states/eV upon inclusion of the SOC).  
Moreover, we have found that SOC effects do not modify significantly neither the structural phase diagram, nor the phonon frequencies at the $\Gamma$-point, and can thus be safely ignored in phonon and electron-phonon calculations.

\begin{figure}
\centering
\includegraphics[width=1\linewidth]{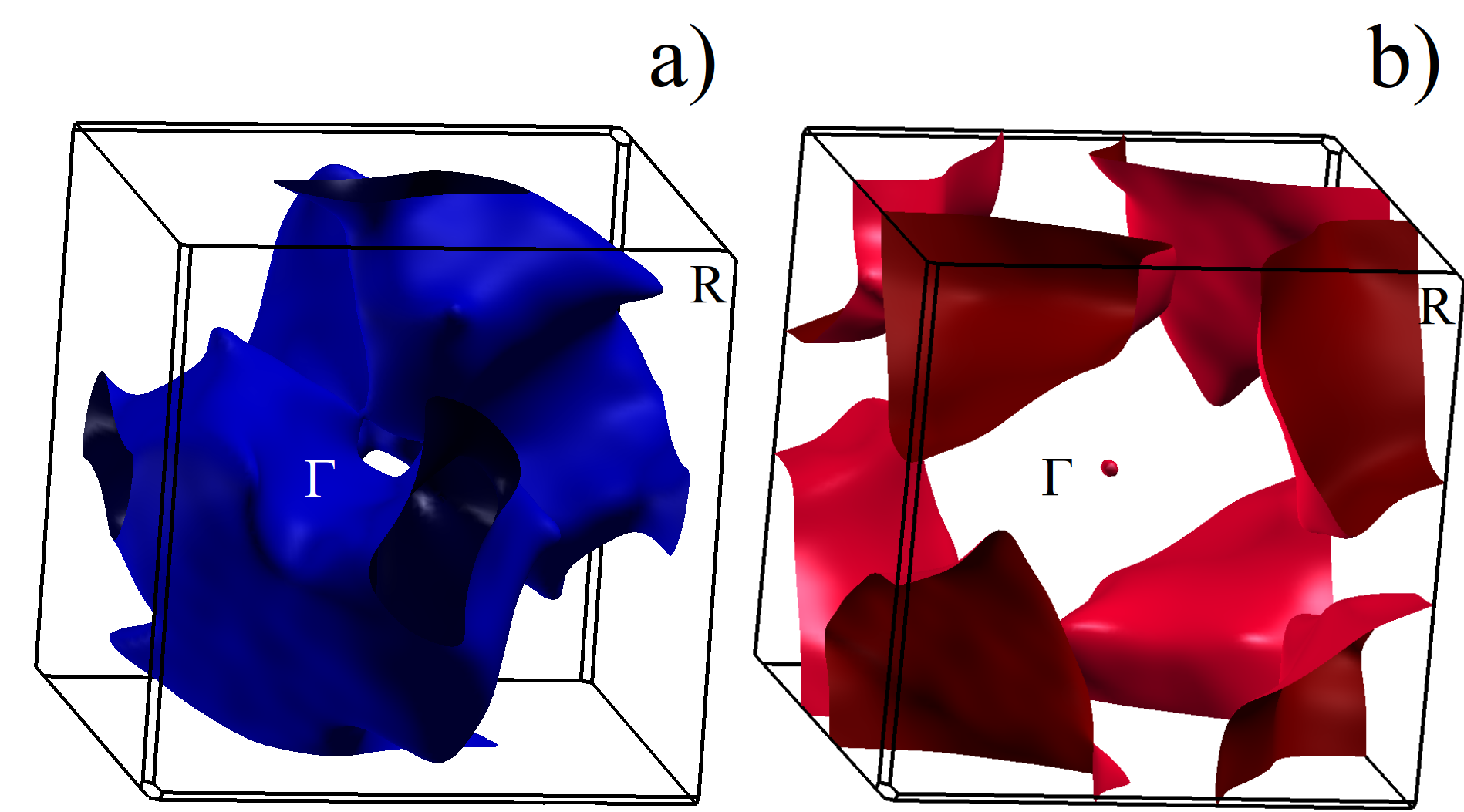}
\caption{Fermi surface of PbMo$_{6}$S$_{8}$ in the the trigonal R$\overline{3}$ crystal structure at 0 GPa without SOC. In panel a) the structure encircling the $\Gamma$ point, in panel b) the structure encircling the R-point with a small pocket around $\Gamma$.}\label{fig4}
\end{figure}

The FS for the R$\overline{3}$ structure without SOC contribution is shown in Fig.~\ref{fig4}. This is composed of a vanishing small pocket around the $\Gamma$ point and two extended sheets, all stemming from the two E$_{g}$ bands. One sheet is a warped structure around the $\Gamma$ point, while the other forms a closed pocket around the corner of the BZ ($R$ point). 

By increasing the external pressure, the bandwidth of the E$_{g}$ states is enhanced, and the total DOS at the Fermi energy reduced (dropping from 10.1 to 7.4 states/eV at 5 GPa). This trend seems to be common to many Chevrel phases\cite{Pea2015, Fischer1978}, and will be crucial to interpret the evolution of the superconducting critical temperature with pressure (see below). 
 
\subsection{Dynamical properties}
\label{sec:phonons}

 \begin{figure}
\centering
\includegraphics[width=1\linewidth]{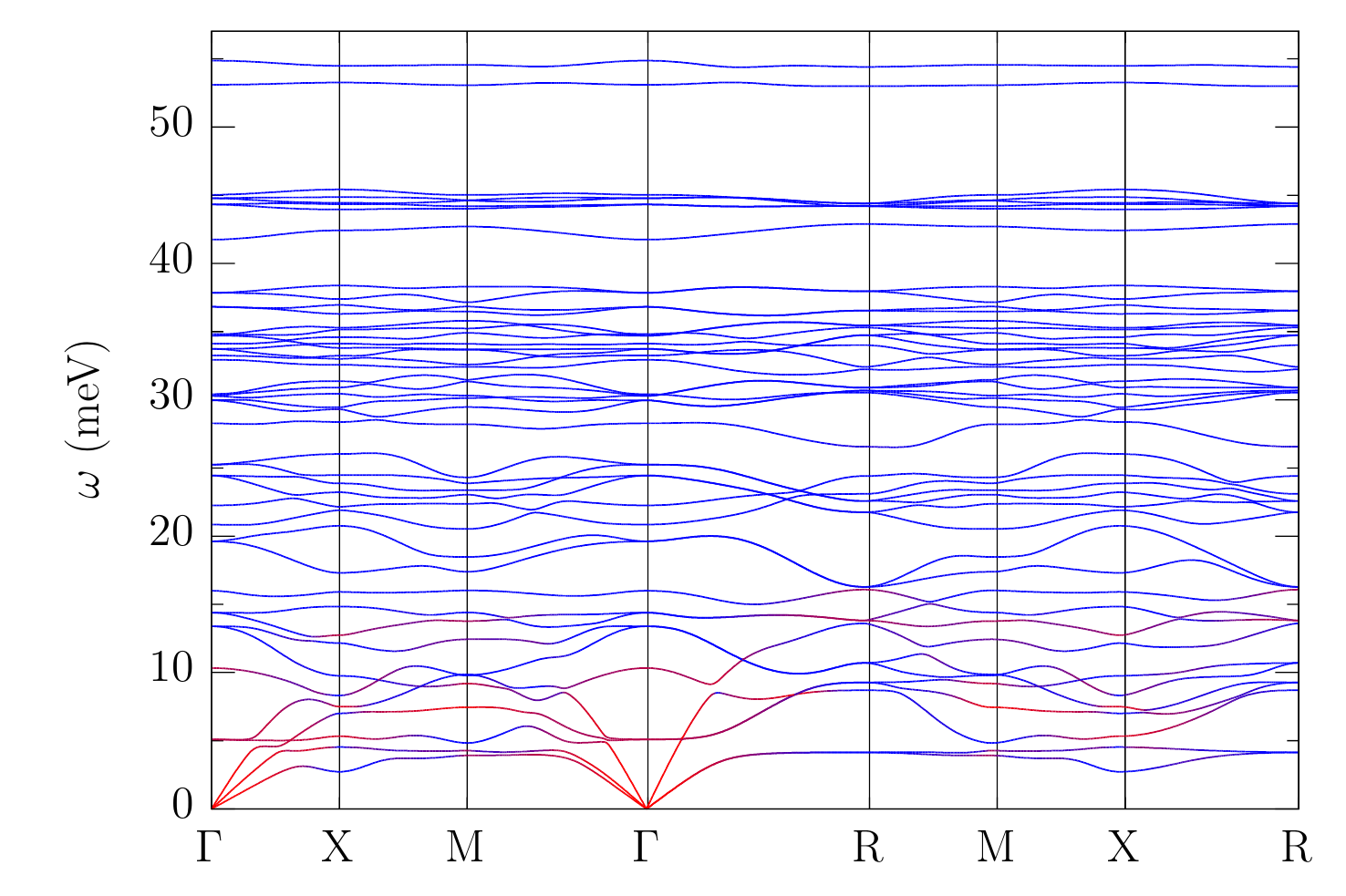}
\caption{Phonon band structure of PbMo$_{6}$S$_{8}$. 
Seeing the lattice as a set of compenetrating molecules, the color code 
(red to blue) 
emphasizes the gradual evolution from  more inter-molecular (red) to more intra-molecular (blue) character of the lattice vibrations.}\label{fig:PHBANDS}
\end{figure}

\begin{figure}
\centering
\includegraphics[width=1\linewidth]{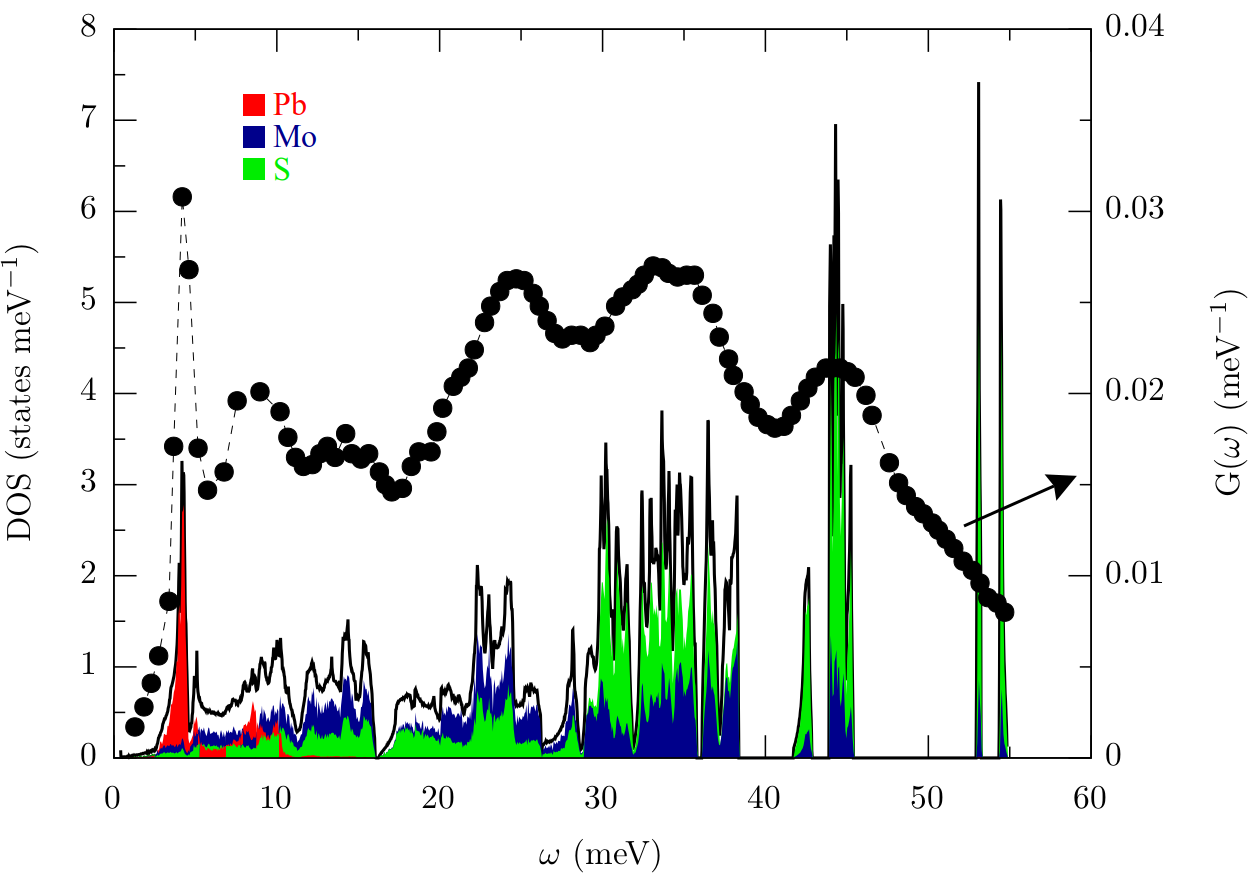}
\caption{Calculated phonon density of states (black line - left axis) compared to the measured weighted phonon spectrum G($\omega$) from Ref.~\onlinecite{Bader1976} (black dots - right axis). The colored areas give atom resolved decomposition of the phonon density of states.}\label{fig:PHDOS}
\end{figure}

We have calculated the phonon dispersion for the relaxed structures at P=0, 1, 3 and 5 GPa. All the spectra show real phonon frequencies, confirming the dynamical stability of the R$\overline{3}$ structure over the considered pressure range. 
Here, we focus our analysis on the simulations at zero pressure, for which we present the phonon dispersion (Fig.~\ref{fig:PHBANDS}) and density of states (Fig.~\ref{fig:PHDOS}). The phonon DOS has been decomposed into the respective atomic components of the vibrational eigenstates. This decomposition highlights a peculiar feature of the lattice dynamics of metal Chevrel phases, that is the presence of a dispersionless Einstein mode caused by the rattling oscillations of the metal (Pb) in its cage. This Pb mode is located at about 4 meV and  is visible as a red peak in Fig.~\ref{fig:PHDOS}. Pb modes dominate the spectrum below 10~meV, giving no contribution above this threshold. Between 10 to 40~meV we observe a mixed Mo-S character of the lattice vibrations, whereas high energy peaks (at 45 and 55 meV) are predominantly of S type. 

The ``molecular-like" crystal structure of Chevrel allows one to qualitatively classify the phonon modes according to their intra- or inter-molecular character\cite{Reichman}.
In order to distinguish between inter- and intra-molecular modes, we have calculated the average cosine of the angle between the atomic displacements for each vibrational mode. This is defined as 
$\alpha^{\mathbf{q},\nu}= \sum_{i,j}Re(\mathbf{v}^{\mathbf{q},\nu}_i \cdot \mathbf{v}^{\mathbf{q},\nu}_j)/|\mathbf{v}^{\mathbf{q},\nu}_i||\mathbf{v}^{\mathbf{q},\nu}_j|$, where $\mathbf{v}^{\mathbf{q},\nu}_i$ is the projection on atom $i$ of the (complex) phonon eigenvector of momentum $\mathbf{q}$ and branch $\nu$. Inter-molecular vibrations, involving the rigid displacement of the cubic cage, are characterized by large values of the average cosine, while smaller values of $\alpha^{\mathbf{q},\nu}$ identify intra-molecular modes.The results of this analysis are shown in Fig.~\ref{fig:PHBANDS} using a color scale. We observe that lower energy modes (up to 17 meV) have inter-molecular character, with the only exception of the aforementioned Pb rattling mode. These modes are separated by a small gap from an intra-molecular region, which comprises Mo-derived modes and high energy (up to 55 meV) S-derived modes\cite{Bader1976,Fischer1978}.

Our theoretical results can be directly compared to the weighted phonon spectrum G($\omega$) measured by Bader {\em et al.}~\cite{Bader1976}, which is also shown in Fig.~\ref{fig:PHDOS}. 
The agreement is remarkable: the calculated DOS well reproduces all the main spectral features of the experimental G($\omega$). In particular, it confirms the presence of a peak at 4.3meV stemming from the doubly degenerate Einstein mode of the Pb atom\cite{Bader1976}, which is indeed missing in the phonon spectrum of pristine Mo$_{6}$S$_{8}$ \cite{Bader1976_book}.

Upon applied pressure (up to 5 GPa), the properties of the lattice dynamics and its mode decomposition remain similar to the P=0 case. The main effect of the pressure is an overall stiffening of the vibrational modes as the system moves away from the R$\overline{3}$-P$\overline{1}$ phase transition point (Fig.~\ref{fig:Enthalpy}). 

\subsection{Electron-phonon coupling}

\begin{figure}[t]
\centering\includegraphics[width=1\linewidth]{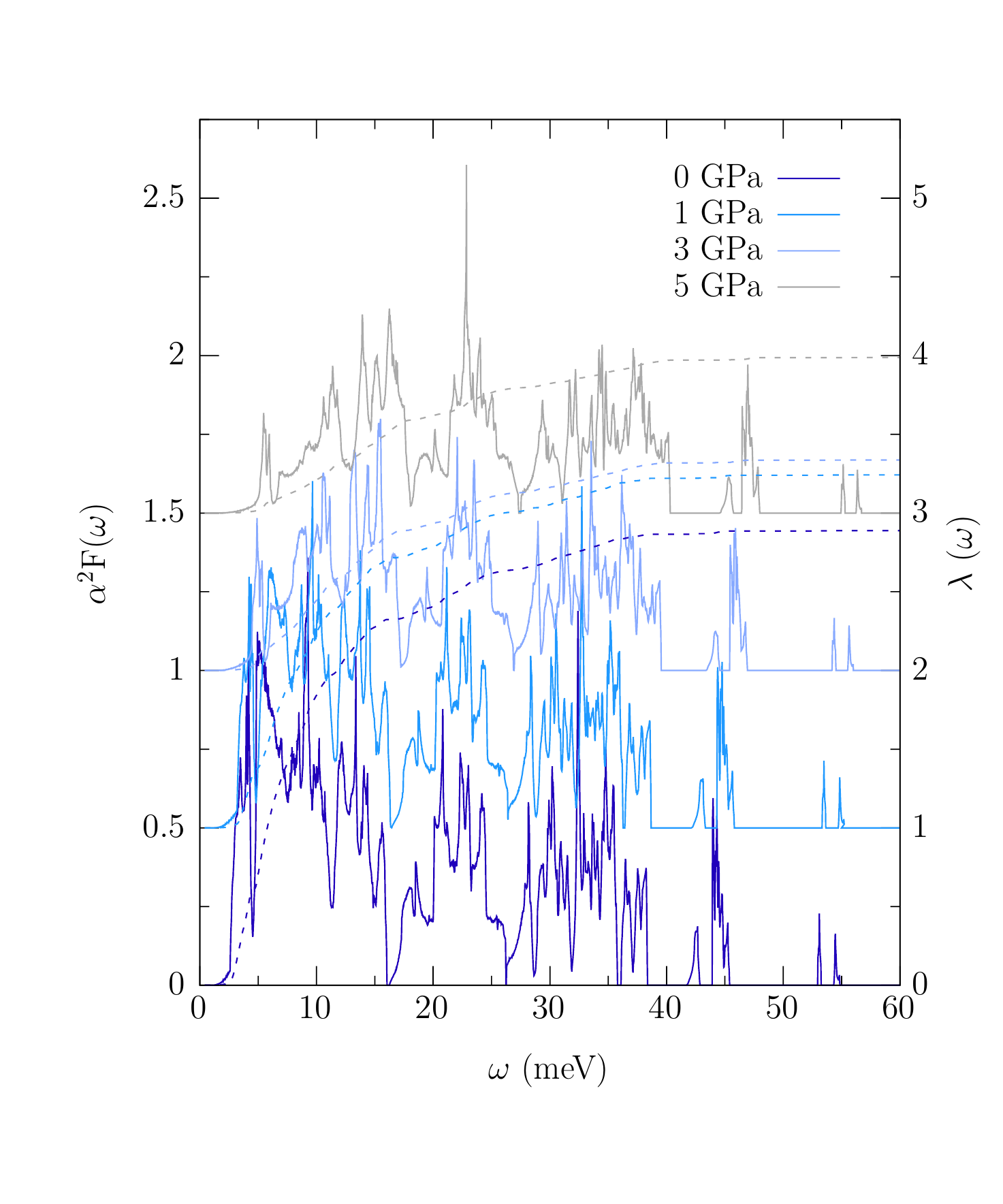}

\vspace{-1cm}

\caption{Pressure evolution of the Eliashberg function $\alpha^{2}F(\omega)$ (full line) and integration curve (dashed line) of the electron phonon coupling parameter $\lambda$. A vertical shift of 0.5 along the vertical axis is applied between consecutive curves to  improve readability.}\label{fig:Eliashberg}
\end{figure}

The stiffening of the lattice vibrations under pressure has a profound effect on the EPC, which strongly depends on the vibrational eigenvalues. The Fermi surface average of the EPC, described by the Eliashberg function of Eq.~\ref{eq:alpha}, is shown in Fig.~\ref{fig:Eliashberg}. This function rapidly loses spectral weight under pressure, so that the integrated coupling parameter $\lambda$ (Eq.~\ref{eq:lambda}) drops from the extremely large value of 2.87 at P=0, to the medium/weak value of 0.98 at 5~GPa.

Our P=0 estimate of $\lambda$ is in agreement with what recently reported in Ref.~\onlinecite{Reichman}. On the other hand,  
a much lower value of $\lambda$, equal to 1.2, has been historically attributed to PbMo$_{6}$S$_{8}$\cite{Fischer1978,Pobell1982}. That value had been obtained from the experimental T$_{c}$ by inverting the McMillan formula~\cite{McMillan_PR1968,AllenDynes_PRB1975}, where an average phonon frequency $\omega_{ln}$ of 12~meV~\cite{Bader1976,Bader1978,Fischer1978} and a conventional value of $\mu^{*} = 0.1$ had been assumed.
Such a strong disagreement between \textit{ab initio} coupling calculations and the Allen-Dynes semi-empirical formula is uncommon and suggests an anomaly in the superconducting behavior of Chevrel phases.

A direct experimental determination of $\lambda$ in Chevrel phases would clarify the issue, however this appears to be a challenging task. Indeed, the complexity of the phonon spectrum and the effects of the variable stoichiometry make difficult an estimation from specific heat measurements~\cite{Pobell1982}. The experimental determination of the electronic density of states itself, N(E$_{F}$), poses problems for its reliability. 
Susceptibility data suggested  
N(E$_{F}$)=18.52 states/eV and $\lambda$ = 1.3, by assuming a (non-interacting) Pauli susceptibility~\cite{Andersen_ElectronicStructureChevrel_PRB1978}.  

On the other hand, including the effective exchange and correlation kernel for the Mo metal, changes the previous values to N(E$_{F}$)$\approx$10.05 states/eV and $\lambda$=2.5~\cite{Andersen_ElectronicStructureChevrel_PRB1978}, in close agreement with our estimates. We also note that, due to the relatively narrow bandwidth of the E$_{g}$ bands, it is not excluded that subtler non-adiabatic effects in the susceptibility might play a role~\cite{Cappelluti2001}.

\begin{table}
\centering
\medskip
\begin{tabular}{cccccc}
\hline
P (GPa) &~\;\; $\lambda$\;\;~ &~\;\; $ ~\;\; \mu \;\;~ $ & T$_{c}^{\!A\!D}$(K) & T$_c^{\!S\!C\!D\!F\!T}$(K) & $\Delta$(meV) \\
\hline
0  &2.87 & 0.49 & 21.85 & 15.1 & 2.8 \\
1  &2.26 & 0.44 & 20.16 & 14.22 & 2.56 \\
3  &1.28 & 0.38 & 14.31 & 10.43 & 1.70 \\
5  &0.98 & 0.34 & 10.57 & 8.22 & 1.26\\ 
\hline
\end{tabular}
\caption{Collected values of the electron-phonon coupling $\lambda$,   Coulomb interaction at the Fermi level $\mu$, superconducting critical temperatures T$_{c}$ (obtained from the Allen-Dynes formula and with SCDFT) and $\Delta$, the SCDFT superconducting gap in the low temperature limit.}\label{tab:data}
\end{table}

\subsection{Superconducting transition}\label{sec:superconductivity}

The most commonly adopted approach to estimate the superconducting critical temperature from first principles simulations is the modified Allen-Dynes formula~\cite{AllenDynes_PRB1975,McMillan_PR1968} in Eq.~\ref{eq:AllenDynesTc}. Apart from electron-phonon parameters, which can be directly extracted~\cite{AllenMitrovic_TheoryofSuperconductingTc_1983} from the Eliashberg functions in Fig.~\ref{fig:Eliashberg}, this equation depends on the Coulomb pseudopotential $\mu^{*}$.
By assuming a rather ``conventional" value\cite{ScalapinoSchriefferWilkins_StongCouplingSC_PR1966,AllenMitrovic_TheoryofSuperconductingTc_1983,CarbotteRPM1990} $\mu^{*}$=0.14, we have obtained a T$_c$ of 21.8~K, which overestimates the experimental value of 14.5~K by $\sim$50\%. 
This discrepancy is much larger than the usual accuracy of the formula\cite{CarbotteRPM1990} 
and has been  already pointed out in Ref.~\onlinecite{Reichman}, where, despite using an anomalously large  $\mu^{*}$ = 0.24, the authors obtain a critical temperature of 18.8 K, still higher than the experimental result.
Such anomalies in Allen-Dynes predictions are often signals of additional interactions/effects which compete with the electron-phonon coupling in bringing about the superconducting condensation. 
Well-known examples are  spin-fluctuations\cite{vanLoon2018,BerkSchrieffer_SF_PRL1966}, anharmonic phonon effects\cite{PhysRevLett.111.177002}, $d$-wave gap symmetry\cite{PhysRevB.75.104501}, anisotropic effects\cite{PhysRevB.73.144512}, vertex corrections\cite{pietronero} and strong correlations effects\cite{Capone2364,PhysRevLett.110.216405}. Surely, this severe overestimation of the T$_c$ hints at the non-conventional (or, rather, anomalous) behavior of Chevrel phases in comparison to straight BCS superconductors.

\begin{figure}
\centering\includegraphics[width=1\linewidth]{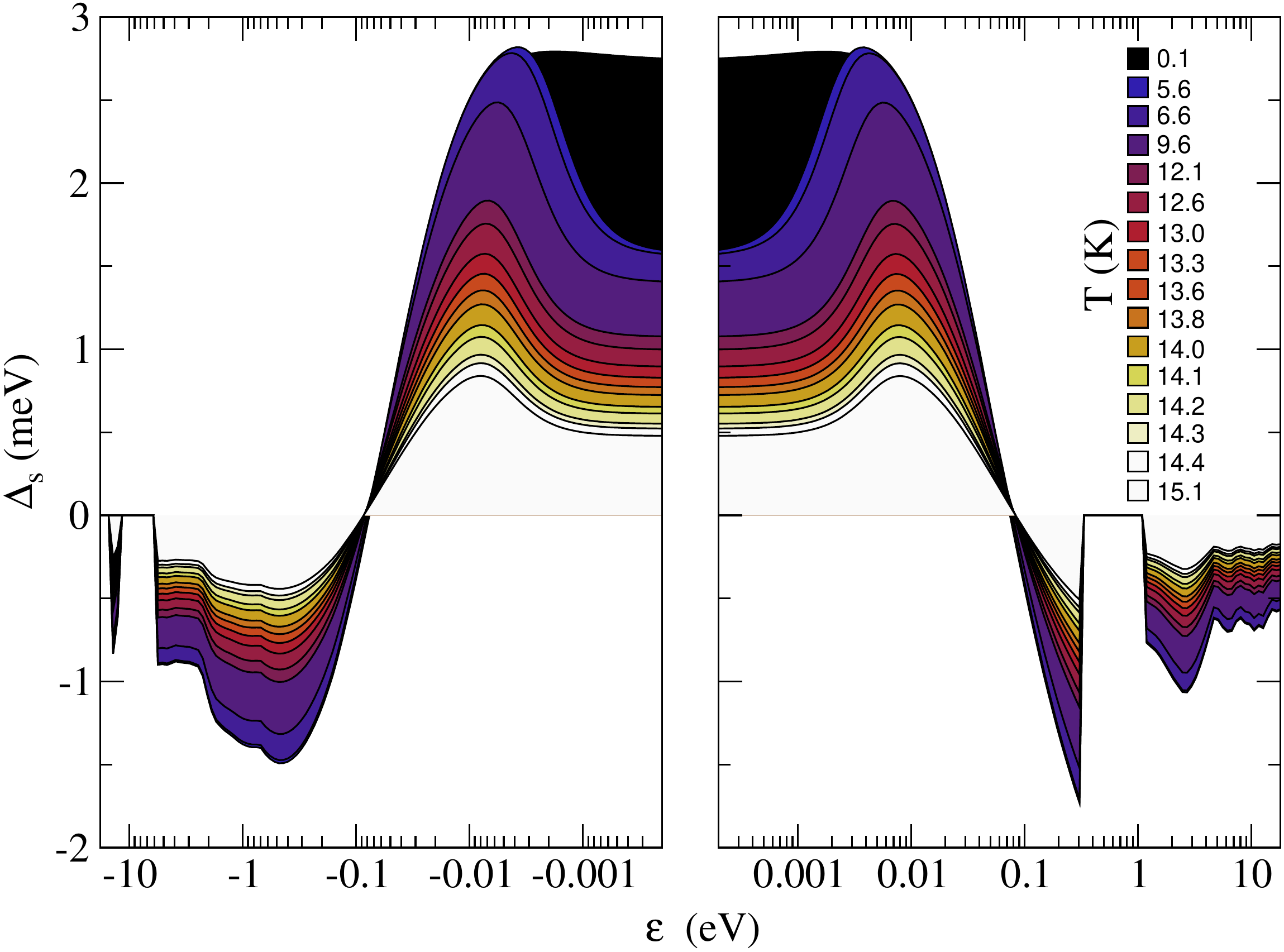}
\caption{SCDFT Kohn-Sham gap obtained by solving Eq.~\ref{eq:SCDFTgapeq} at P=0 and as a function of temperature. The gap is sharply peaked near the Fermi level (note the semi-logarithmic scale). Negative values of the gap indicate the effect of Coulomb interactions, which cause a phase shift of $\pi$ in the gap function, occurring above the phononic energy scale. 
Above the Fermi level the Coulomb contribution is strongly reduced by the presence of a large band gap, which reduces the efficiency of the Coulomb renormalization mechanism.} 
\label{fig:KSgap}
\end{figure}

However, one should bare in mind that the treatment of Coulomb effects within the conventional Eliashberg approach~\cite{AllenMitrovic_TheoryofSuperconductingTc_1983} is often too crude, and a proper inclusion of the Coulomb interaction from first-principles may be needed to handle the delicate balance between phonon-mediated attraction and Coulomb repulsion\cite{Sanna_GenuinePredictions_JPSJ2018,Davydov_AbinitioEliashberg_arxiv2020}. Superconducting density functional theory was formally developed along this line and has now reached predictive accuracy for conventional superconductors\cite{OGK_SCDFT_PRL1988, Lueders2005,Marques2005,Sanna_SCDFT2017,FLORESLIVAS20201}.
In this work we have solved the self-consistent SCDFT-KS gap equation~\ref{eq:SCDFTgapeq} as a function of temperature and pressure by including the electron-phonon interaction in the SPG approximation and the Coulomb repulsion in static RPA. The temperature-dependence of the resulting KS gap, $\Delta_s$, at P=0 GPa is shown in Fig.~\ref{fig:KSgap}.

\begin{figure}
\centering\includegraphics[width=1\linewidth]{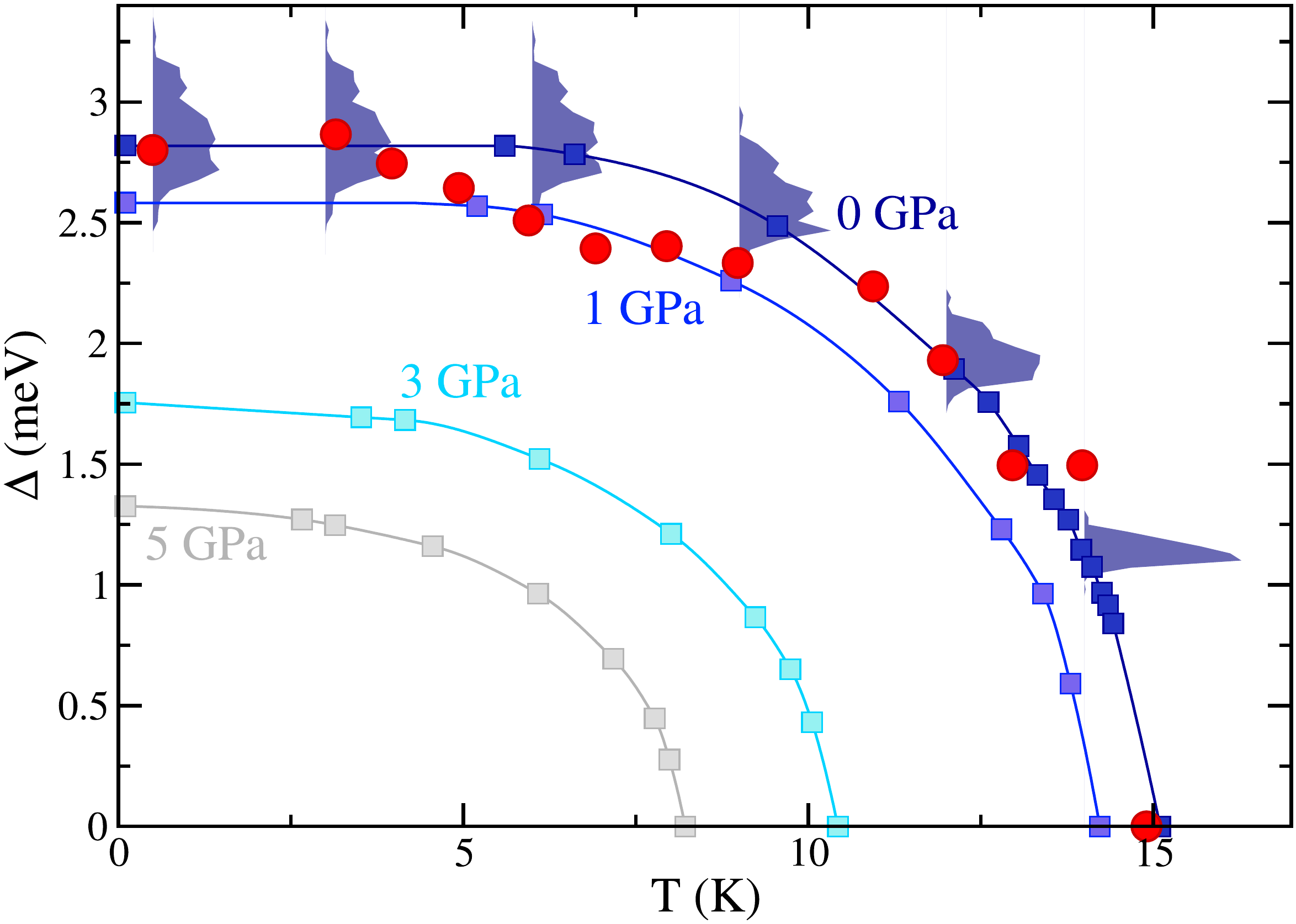}
\caption{Superconducting energy gap as a function of pressure and temperature, as obtained in SCDFT (squares) compared to the experimental data from Ref.~\onlinecite{Petrovi2011} (red circles). Shades indicate the anisotropic gap distributions at 0 GPa).}\label{fig:GapTvsEXP}
\end{figure}

The SCDFT values of the physical superconducting gap $\Delta(T)$ at different pressures are presented in Fig.~\ref{fig:GapTvsEXP}. Here, we have included a comparison to the experimental estimates (at ambient pressure) from the tunneling measurements by Petrovi\`c and coworkers\cite{Petrovi2011}.

The calculated critical temperature at P=0 GPa is T$^{\!S\!C\!D\!F\!T}_c$=15.1 K, to be compared with an experimental value of $\simeq$ 14.5 K (equal to the average of the available experimental data \cite{Petrovi2011,Shelton1975,Matthias1972,Fischer1978}).

The typical errorbar of SCDFT, which, we emphasize, is an \emph{ab initio} approach completely free from adjustable parameters, is of the order of 15\% of Tc~\cite{SPG_EliashbergSCDFT_PRL2020}.
Therefore, given the complexity of the experimental samples due to non-ideal stoichiometry and coexistance between multiple structural phases, the accuracy of the predicted T$_c$ is excellent. 
In particular, our result demonstrates that an \emph{ab initio} treatment of the  Coulomb interaction is of crucial importance for the correct description of superconductivity in Chevrel phases. Our SCDFT estimate for the zero-temperature gap at ambient pressure, $\Delta(0)$ = 2.79 meV, is also in good agreement with the reported experimental value of 3.06 $\pm$ 0.15 meV\cite{Petrovi2011}. 
We observe that the temperature dependence of the superconducting gap $\Delta(T)$, as obtained in SCDFT, has a BCS-like behavior at all the pressure values considered in Fig.~\ref{fig:GapTvsEXP} and, notably, closely reproduces the experimental curve at P=0 GPa. The only missing feature is a kink of unknown origin in the experimental points (in between 4 and 10~K). To the best of our knowledge this anomalous feature of the spectrum was first reported by Dubois \textit{et al.} in Ref.~\onlinecite{PhysRevB.75.104501}, but not discussed. Nevertheless, based on a fit of the tunneling spectra by $s$- and $d$-wave models, this work suggested a $d$-wave symmetry of the superconducting gap for the Pb Chevrel phase.
The same kink was later reported by some of the authors attributing its origin to a two-band superconducting phase arising from the two E$_{g}$ bands crossing the Fermi level\cite{Petrovi2011}.
Although these hypotheses are suggestive
we believe that further measurements are probably needed in order to clarify and/or analyze the actual behavior of the superconducting gap in the aforementioned range of temperatures. We observe that the SCDFT gap equation with the kernels here employed does not admit a $d-$wave solution, and even if this phase could be stabilized, owing to the
extremely large electron-phonon strength, it would
 not give a critical temperature in the experimental range.

On the other hand, the possibility of a two-gap superconductivity has been recently addressed by first-principles methods\cite{Reichman}. However, although largely overestimating the T$_c$, these calculations do not support a two-gap scenario. To confirm this indication we have solved (at P=0) the \textbf{k}-resolved SCDFT gap equation, simulating the perfect clean limit of the superconducting state. The obtained results are presented in Fig.~\ref{fig:GapTvsEXP} in the form of a gap histogram at selected temperatures and show that the anisotropy of the  superconducting gap is weak (of the order of 10\%).  The gap distribution is centered at the isotropic value without any evidence of a multigap superconducting state. Indeed, the predicted critical temperature is the same (within numerical accuracy) as for the isotropic case. 
We also observe that the kink in the experimental $\Delta(T)$ reported in Ref.~\onlinecite{Petrovi2011} is not an anisotropic feature of the superconducting gap, which suggests the importance of an experimental reassessment of the temperature dependence of the superconducting gap.

\begin{figure}
\includegraphics[width=1\linewidth]{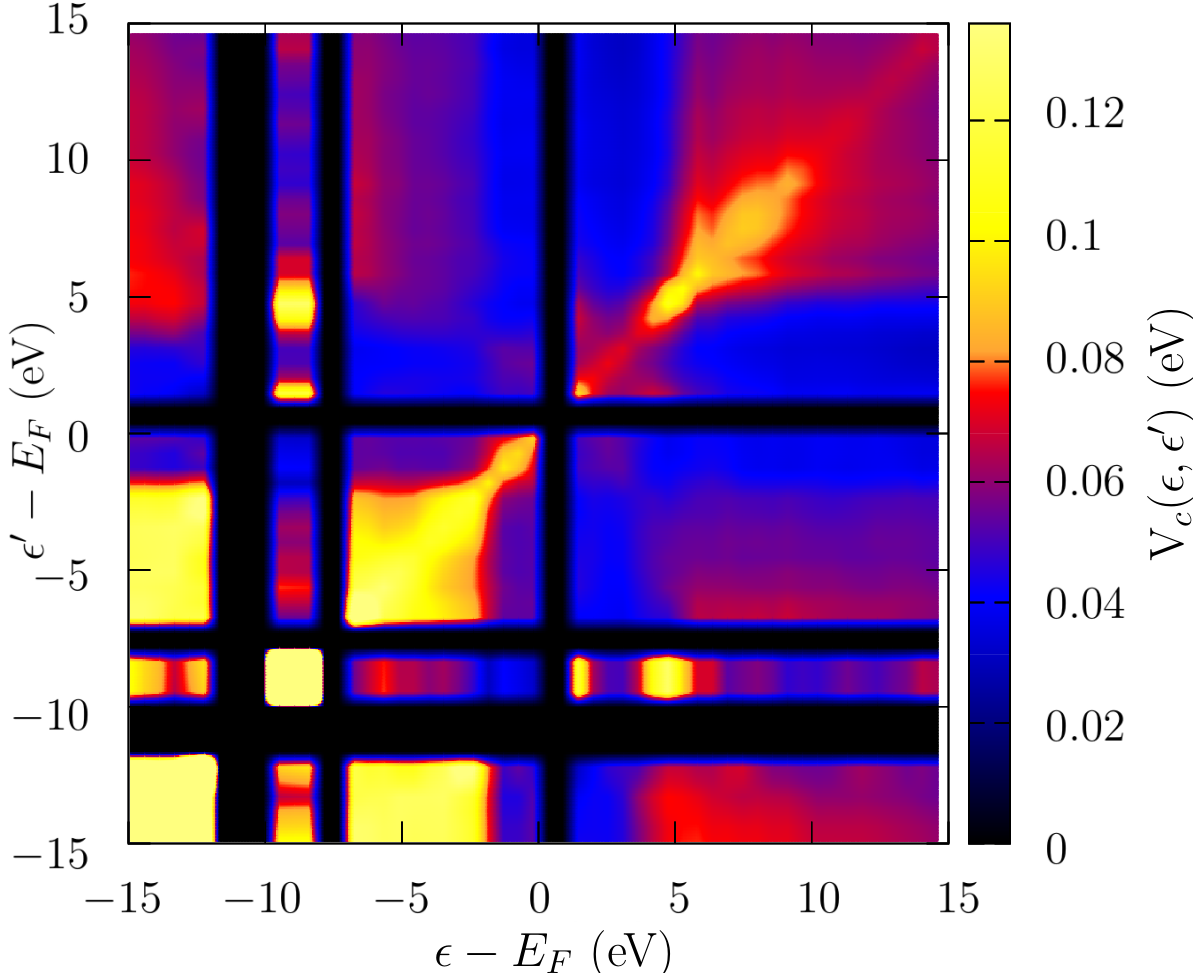}
\includegraphics[width=1\linewidth]{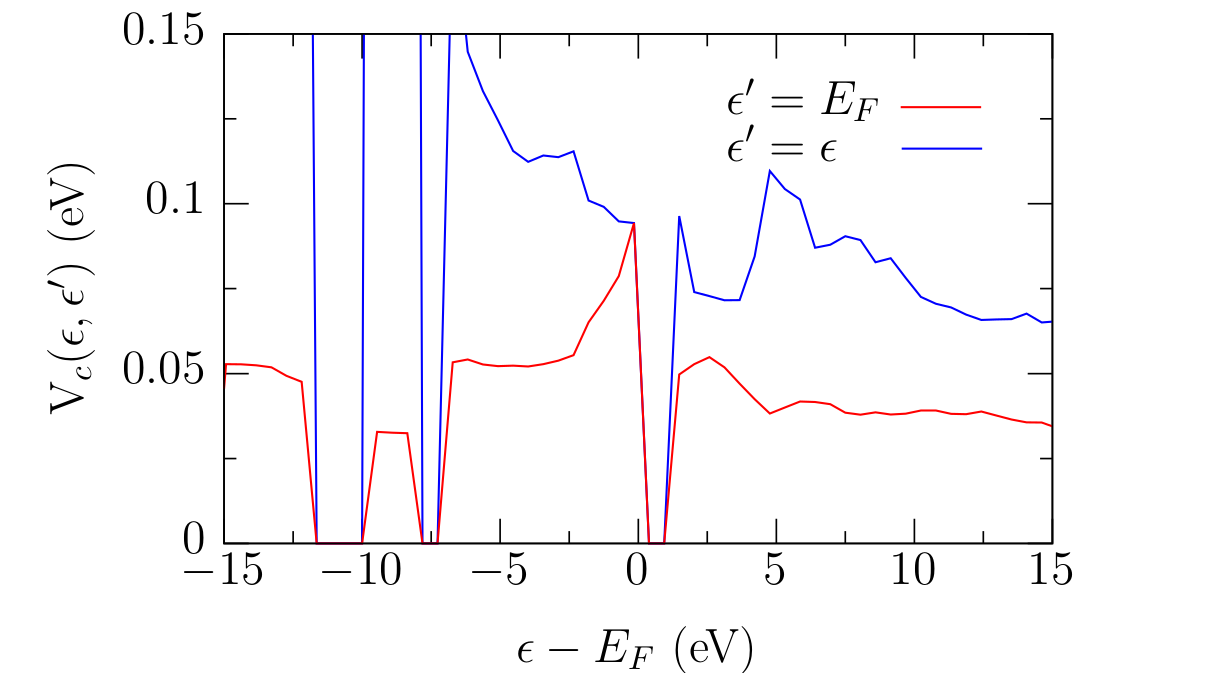}
\caption{Screened Coulomb interaction $V_c$  computed in the random phase approximation and averaged over iso-energy surfaces ($\epsilon,\epsilon'$). 
The top panel shows the entire function using a color-scale (black to yellow). The bottom panel shows two cuts of $V_c$, the diagonal (blue) and the cut relative to the Fermi level (red). The latter is the most important for superconductivity since it describes how strongly states at the Fermi level scatter with states in the entire valence and conduction region.}
\label{fig:CoulombKernel}
\end{figure}

The reason why SCDFT is able to achieve an accurate description of the superconducting state in terms of gap and $T_c$ is, as already mentioned, the fully \emph{ab initio} inclusion of Coulomb interactions. The remarkable difference between the Allen-Dynes and SCDFT results can be traced back to uncommon features of the Coulomb repulsion in Chevrel phases. 

The Coulomb parameter, $\mu$=0.49 at P=0 GPa, is larger than the typical values calculated for elemental metals\cite{Davydov_AbinitioEliashberg_arxiv2020,Sanna_CaC6_PRB2007}, and only comparable to anomalous cases such as palladium (0.977), platinum (0.798), scandium (0.523) and vanadium (0.512)\cite{kawamura2020}). 
This large Coulomb parameter can be partially ascribed to the large N($E_{F}$)=10.1 states/eV entering Eq.~\ref{mu}. However, when compared to other elements with the same N(E$_{F}$) (as listed in Ref.~\onlinecite{kawamura2020}), PbMo$_{6}$S$_{8}$ turns out to have higher $\mu$, which is, thus, indication of  sizable Coulomb matrix elements at the Fermi level.

It should be noted that a large value of $\mu$ does not necessarily imply a low T$_c$. The reason is that Coulomb interactions in superconductors occur on two distinct energy scales. Within the phononic (Debye) energy scale, Coulomb interactions are strongly repulsive and tend to reduce the Cooper pair binding; on the large electronic (Fermi) energy scale, instead, they effectively act as attractive forces, in a mechanism that is commonly known as Coulomb renormalization~\cite{MorelAnderson_PR1962,ScalapinoSchriefferWilkins_StongCouplingSC_PR1966}. It follows that a strong Coulomb repulsion at the Fermi energy is not meaningful per se, since it can be compensated by an equally strong repulsion between states at the Fermi level and other states in the valence/conduction region~\cite{ScalapinoSchriefferWilkins_StongCouplingSC_PR1966}. 

The actual effect of Coulomb interactions, thus, is not merely described by the value of $\mu$, but is hidden in the energy dependence of the screened Coulomb matrix elements $V_c(\epsilon,\epsilon')$ (Fig.~\ref{fig:CoulombKernel}), the electronic DOS (Fig.~\ref{fig:elBANDSandDOS}) and the self-consistent solution of the gap equation (Fig.~\ref{fig:KSgap}).

\begin{figure}
\centering\includegraphics[width=1\linewidth]{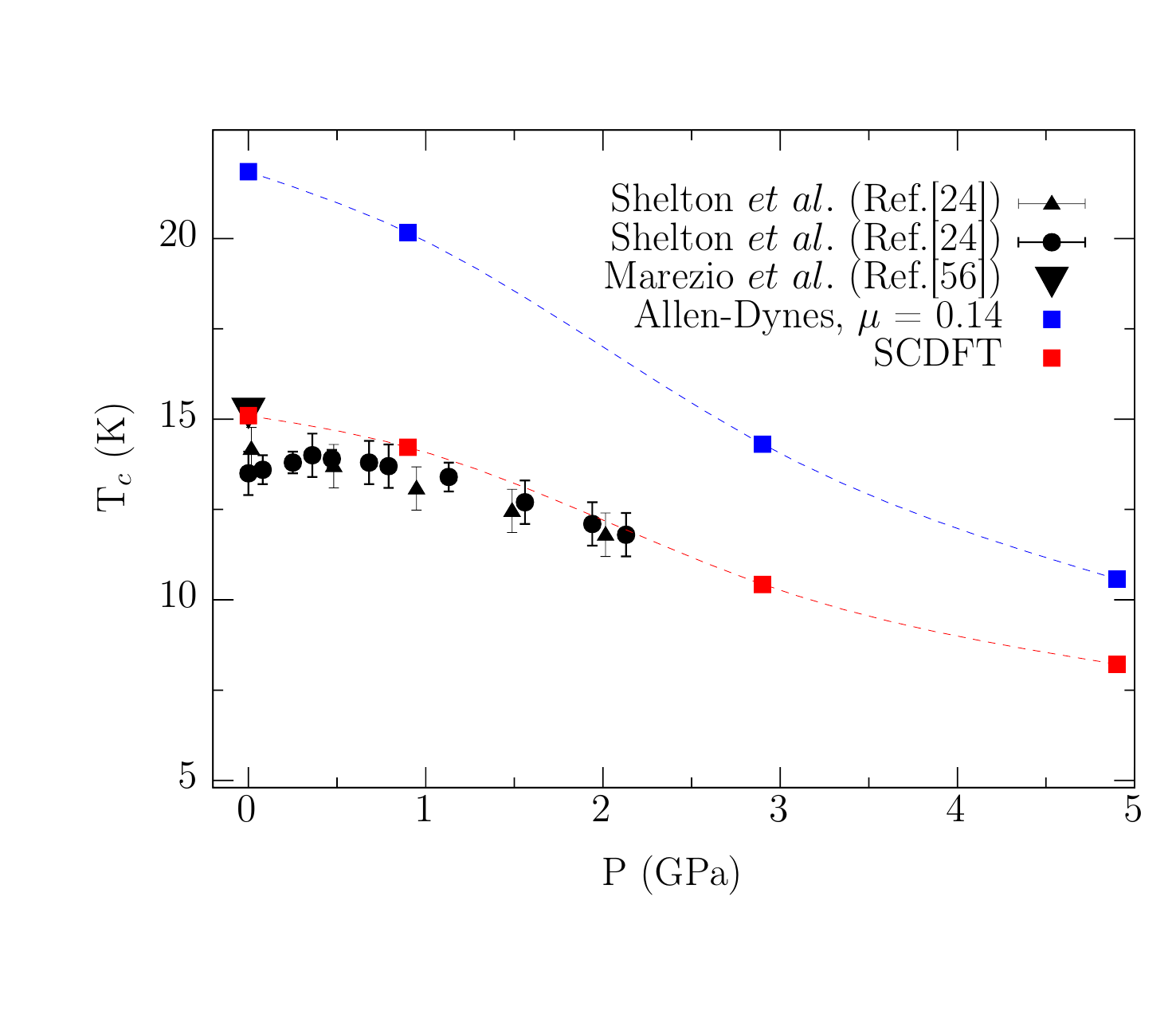}
\caption{Superconducting critical temperature T$_{c}$ as a function of pressure, compared to the experimental data\cite{Shelton1975,Marezio1973} in literature.}\label{fig:SCDFTvsEXP}
\end{figure}

The main peculiarity of PbMo$_{6}$S$_{8}$ lies in the energy dependence of the screened Coulomb interaction. As one can see in the upper panel of Fig.~\ref{fig:CoulombKernel}, $V_c(\epsilon,\epsilon')$ has a sharp peak at the Fermi level (yellow peak at (0,0)), but rapidly decreases when moving away from it (blue ``cross-like" shape around (0,0)). This means that the Coulomb repulsion is strong between states at the Fermi level, but gets considerably weaker when one of the two interacting electrons is far from Fermi.

This behavior is also apparent when looking at an horizontal cut of the 2D plot (red curve in the bottom panel of Fig.~\ref{fig:CoulombKernel}), which shows a peak near $\epsilon=E_F$. An other uncommon feature of PbMo$_{6}$S$_{8}$, with respect to conventional metals, is the presence of a wide band gap above the Fermi level (black bands in the upper panel of Fig.~\ref{fig:CoulombKernel}), which essentially removes a large fraction of the conduction region. The combination of these two anomalies results in the little effectiveness of the Coulomb renormalization mechanism. This can be proved by two simple numerical tests:
If we assume in SCDFT a constant energy approximation for the Coulomb repulsion by retaining its value at E$_{F}$\cite{ScalapinoSchriefferWilkins_StongCouplingSC_PR1966}, while keeping the full energy-dependence in the density of states, i.e., if we use $\mathcal{K}^c_{\epsilon,\epsilon'}=N(\epsilon)V_{c}\left(E_F,E_F\right)$ in Eq.~\ref{eq:SCDFTgapeq}, the resulting T$_{c}$ (at 0 GPa) is 18.4 K. In fact, we are enhancing the Coulomb renormalization at high energies by attributing to all the electrons the same (strong) Coulomb repulsion. This overestimation of the experimental T$_{c}$ is in line with the results obtained in Ref.~\onlinecite{Reichman}, where the authors adopted a conceptually similar approximation (here the energy dependence of the electronic DOS was effectively captured by using a low value of the Fermi energy in the Morel-Anderson formula).
If we assume, in addition, a constant density of states, $N(\epsilon)=N(E_F)$, then SCDFT predicts a critical temperature of the order of 24~K, close to that provided by the Allen-Dynes formula with the conventional $\mu^*$.
These two computational experiments clarify the anomalous critical temperature of PbMo$_{6}$S$_{8}$ despite the large electron-phonon coupling.

Fig.~\ref{fig:SCDFTvsEXP} shows the pressure dependence of the superconducting T$_{c}$,  compared to the experimental T$_{c}$(P) curves. SCDFT predictions are accurate for all the investigated pressures, whereas the Allen-Dynes approach gives a systematic overestimation, indicating that the Coulomb interaction acts in a similar manner at low and high pressure.
Increasing pressure has a detrimental effect on the superconducting properties of PbMo$_{6}$S$_{8}$ because it lowers the Eliashberg $\alpha^2F$ function of Eq.~\ref{eq:alpha} by reducing N(E$_{F}$) and by stiffening the low energy modes ($|g_{\textbf{k},\textbf{k}'}^{\nu, m n}|^2 \propto \omega_{{\bf q}\nu}^{-1}$).
However, pressure effects on $\alpha^2F$ are partially mitigated by a weakening of the Coulomb interaction at the Fermi level (see Tab.\ref{tab:data}), which follows the reduction in the electronic DOS according to Eq.~\ref{eq:Kc}.

\section{Conclusions}
We have shown that joint first-principles DFT and SCDFT calculations are able to accurately 
describe the electronic,  
structural and superconducting properties of Chevrel phases, which had previously remained partly unexplained. 
Taking PbMo$_{6}$S$_{8}$ as a prototype material, we have found that the low energy electronic structure around the Fermi energy is determined by the molybdenum $d_{x^2-y^2}$ states in a E$_{g}$ symmetry, on which the spin-orbit interaction induces a topological change of the Fermi surface 
-- a small additional electron pocket near $\Gamma$ --
 in principle experimentally verifiable.
The structurally determined phonon spectrum exhibits
a characteristic Einstein Pb-derived mode and a 
strong energy  separation between high-energy intra-molecular and low/intermediate inter-molecular modes,  the latter being the most coupled with electrons at the Fermi energy. 
We predict that PbMo$_{6}$S$_{8}$ should be close to a low-energy structural distortion, 
which can be promoted by  chemical and external pressure, and 
whose proximity enhances the electron-phonon coupling.  
On the other hand, we have found at negative pressures no evidence of impending metal-insulator or magnetic transitions, which might suggest the possible neighborhood of Mott
states and strongly correlated superconductivity.

By means of SCDFT we have predicted the superconducting critical temperature, its variation with the external pressure, the superconducting gap and its temperature  evolution, with unprecedented 
accuracy and parameter-free agreement with experiments.

Our simulations point to a peculiar role of the Coulomb repulsion between localized Mo $d$-states at the Fermi energy, which, to the best of our knowledge, is stronger than that of any other phononic superconductor. Nevertheless, the superconducting mechanism in Chevrel compounds is conventional. Their ``unconventional" position in the Uemura plot is indeed misleading and can be explained due to the narrow bandwidth and large electron-phonon effective mass, which yield a very low experimental estimate of the Fermi temperature.

In conclusion, this work opens the possibility for a systematic re-investigation of the interesting family of Chevrel phases, looking at new aspects like the role of the spin-orbit interaction, measurements of the electronic band structure, effects of reduced dimensions
(2D, 1D), exploration of different chemical doping\cite{Lin2020} and studies of related superconductors like K$_2$Cr$_3$As$_3$\cite{Bao_SCK2Cr3As3_PRX2015}.

\acknowledgements 
G. M. and G. P. acknowledges support from CINECA Supercomputing Center through ISCRA project.
G. P. acknowledges financial support from the Italian Ministry for Research and Education through PRIN-2017 project ``Tuning and understanding Quantum phases in 2D materials - Quantum 2D" (IT-MIUR Grant No. 2017Z8TS5B).
A.S. acknowledges hospitality of the Physics Department of \textit{La Sapienza}, under the program ``Professori Visitatori 2020".
E.T. is supported by ERC Advanced Grant N. 834402 ULTRADISS, and by the Italian Ministry of University and Research through PRIN UTFROM N. 20178PZCB5.

\bibliography{bibliography}

\end{document}